# Nanoscale Observation of Alkane Delayering

M. Bai,[1] K. Knorr,[1,*] M. J. Simpson,[1] S. Trogisch,[1,†] H. Taub,[1,††] S. N. Ehrlich,[2] H. Mo,[3]
U. G. Volkmann,[4] F. Y. Hansen[5]

[1]Department of Physics and Astronomy and University of Missouri Research Reactor,
University of Missouri-Columbia, Columbia, Missouri 65211

[2]National Synchrotron Light Source, Brookhaven National Laboratory,
Upton, NY 11973

[3]Department of Physics and Astronomy, Northwestern University, Evanston, IL 60208-3112

[4]Facultad de Física, Pontificia Universidad Católica de Chile, Santiago 22, Chile

[5]Department of Chemistry, Technical University of Denmark, IK 207 DTU,
DK-2800 Lyngby, Denmark

**Abstract**

Noncontact Atomic Force Microscopy and synchrotron x-ray scattering measurements on dotriacontane ($n$-$C_{32}H_{66}$ or C32) films adsorbed on $SiO_2$-coated Si(100) wafers reveal a narrow temperature range near the bulk C32 melting point $T_b$ in which a monolayer phase of C32 molecules oriented perpendicular to surface is stable. This monolayer phase undergoes a delayering transition to a three-dimensional (3D) fluid phase on heating to just above $T_b$ and to a solid 3D phase on cooling below $T_b$. An equilibrium phase diagram provides a useful framework for interpreting the unusual spreading and receding of the monolayer observed in transitions to and from the respective 3D phases.

Although a macroscopic theory of the wetting of solid surfaces by liquids has been developed some time ago [1], the general question of how molecular rotational and conformational degrees of freedom may influence the wetting of a liquid film interacting with a solid surface via van der Waals forces has not been investigated extensively and is poorly understood. Normal alkane molecules [$C_nH_{2n+2}$] of intermediate length ($15 < n < 50$) provide model systems with which to address this issue. Torsional motion about C-C bonds in their carbon backbone results in conformational changes of the molecules that play an essential role in determining the structure and melting of alkane monolayers [2]. Furthermore, the effect of these molecular conformational changes on wetting can be investigated systematically as a function of the alkane chain length n. As alkanes are the principal constituents of commercial lubricants [3,4], their wetting properties are also of interest for lubricating nanoscale devices such as computer hard drives and micro-electro-mechanical systems.

It is generally believed that films of shorter alkanes ($n \leq 5$) completely wet a solid substrate, i.e., form an infinitely thick film, at the bulk triple point. This has only recently been demonstrated explicitly for the case of pentane ($n$-$C_5H_{12}$) adsorbed on basal-plane graphite surfaces [5]. Experiments with intermediate-length alkanes ($15 < n < 50$) adsorbed on $SiO_2$ surfaces have not found triple-point wetting. Instead, liquid droplets have been observed by optical reflection microscopy at the bulk melting point [6–8]. Complete wetting of intermediate-length alkane films a few degrees above the bulk melting point was inferred indirectly by x-ray specular reflectivity [6,9] and stray light intensity measurements [10].

In this report, we present direct evidence from noncontact mode Atomic Force Microscopy (AFM) images that a film of an intermediate-length alkane, dotriacontane ($n$-$C_{32}H_{66}$ or C32), does not completely wet a $SiO_2$ surface at any temperature. Instead, we find a narrow temperature range near the C32 bulk melting point $T_b$ where a monolayer phase in which the molecules are oriented perpendicular to the surface is stable. On heating just above $T_b$, this monolayer phase undergoes a delayering transition to three-dimensional (3D) droplets that remain present up to their evaporation point. Moreover, the system shows an unusual reentrant drying behavior in that the monolayer phase also undergoes a delayering transition to a 3D solid phase on cooling below $T_b$.

The AFM measurements on C32 films were performed with a Nanoscope IIIa (Veeco Instruments, Inc.) operating in the "noncontact" or tapping mode. Using silicon cantilevers with a resonance frequency of ~33 kHz, we simultaneously recorded AFM images of topography and phase angle (cantilever oscillation relative to the drive) as a function of temperature. The samples reported on here were made by dip-coating an acid-cleaned, electronic-grade Si(100) substrate in a solution of C32 dissolved in heptane (C7) [11,12]. They typically had native oxide coatings with thickness in the range 12–25 Å [11].

In Fig. 1(a), we see a topographic AFM image of a sample at room temperature taken immediately after dip coating. The image shows an island of adsorbed C32 having a "dragonfly" shape. Topographic cross sections indicate the height of this feature to be ~4.2 nm after calibration using the AFM contact mode [12,13] and x-ray specular reflectivity measurements [11]. Because this height is approximately equal to the all-*trans* length of the C32 molecule, we interpret the dragonfly feature as consisting of a single layer of molecules oriented with their long axis perpendicular to the surface. Hereafter, we refer to such a structure as a "perpendicular monolayer." This feature is qualitatively similar to the fractal-like islands previously observed by AFM at low coverages of C30 on $SiO_2$ [14]. X-ray specular reflectivity [11] and contact mode AFM measurements [12] at room temperature indicate that these perpendicular monolayer islands reside on one to two layers of C32 molecules oriented with their long axis parallel to the $SiO_2$ surface.

After heating this sample above the bulk melting point of C32 at 68 °C and returning to 47 °C, we find a qualitative change in topography. As shown in the first image of Fig. 1(b), a typical scan area no longer contains perpendicular monolayer islands but only mesa-shaped particles that cross sections indicate have a height of 30–40 nm and that x-ray

5diffraction reveals have an orthorhombic structure [11]. We cannot, however, exclude the presence of a small number of metastable monolayer islands out of the scan area (see below). Figure 1(b) follows the evolution of one of the 3D particles in a second heating cycle. As the sample is heated from 47 °C to 58 °C, the particle boundary becomes smoother. At ~61 °C, an abrupt decrease in volume occurs [see left insert] with the image at 62 °C showing the particle with an even smoother boundary. Apparently, the lost material enters an interfacial gas phase by which it can be transported out of the scan area.

Between 64 °C (not shown) and 66 °C in Fig. 1(b), we observe a perpendicular monolayer beginning to spread outward with a height of 4.2 nm at the expense of the volume of the mesa-shaped particle located near its center. A cross section of the image at 66 °C (right insert) shows a lower step corresponding to the spreading perpendicular monolayer and two higher steps of four and three perpendicular layers, respectively, within the particle. The perpendicular monolayer continues to spread until, at ~69 °C, it reaches its maximum lateral extent and the material in the mesa-shaped particle has been nearly exhausted. Holes now develop in the perpendicular monolayer; and, at 70 °C, there is an abrupt transition to 3D droplets identified both by their spherical profile and circular perimeter as well as by a large positive increase in the phase angle measured. In other similarly prepared samples that we investigated to higher temperatures, we observed no wetting of C32 to occur up to a temperature of 85 °C at which there is significant thermal desorption into the 3D gas phase.

On cooling this sample just below the bulk melting point at 68 °C, we see in Fig. 1(c) that a perpendicular monolayer begins to spread outward from one of the droplets as the droplet volume decreases. At 57 °C, the material in the droplet is nearly exhausted and holes begin to appear in the monolayer. With further cooling of the sample to 54 °C, we observe a second perpendicular layer form on top of the first. At this point, the combined volume of the two layers begins to decrease due to loss of material to the interfacial gas phase, and the film kinetics slows. The last three AFM images at 51°C show the subsequent evolution as a function of time. The second perpendicular layer grows in area at the expense of the first layer as their combined volume continues to decrease until the two layers match boundaries to form a bilayer island. In other samples, we have observed growth of such bilayer islands into thicker mesa-shaped particles by transport of C32 from monolayer islands of perpendicular molecules through the interfacial gas phase.

We interpret the sequence of AFM images in Fig. 1(b) as indicating the stability of a monolayer phase of C32 molecules oriented perpendicular to the surface in the temperature range from 64 °C to 69 °C. On heating, the kinetics of the transition from the 3D mesa-shaped particles to the monolayer phase is sufficiently slow that we are able to view the growth of the monolayer phase in time (the AFM images are recorded continuously at a rate of ~4.5 min per image). The transition is driven by the lower chemical potential of the monolayer phase. Similarly, on cooling, the transition from 3D liquid droplets to the perpendicular monolayer phase shown in Fig. 1(c) is kinetically hindered so that we again view the growth of the monolayer as an outward spreading from a 3D droplet that has a higher chemical potential. This spreading behavior appears similar to that previously observed in undercooling experiments near the bulk melting point in C30 films [7].

We have also performed similar heating/cooling cycles on higher coverage samples. For these, we observe spreading of a perpendicular monolayer phase to begin at ~62 °C; however, in this case, the number and size of the source particles are sufficient to allow the perpendicular monolayer phase to fill the entire scan area. The delayering transition to 3D droplets is again initiated by the development of holes in the perpendicular monolayer, but the kinetics of the transition is slow enough that we observe a gradual decrease in the area occupied by the perpendicular monolayer in a temperature range from 69 °C to 74 °C. This behavior is similar to the previously observed recession of a perpendicular monolayer [7] as a C30 film is heated above its bulk melting point.

In Fig. 1(b), we characterize the mesa-shaped particle from which the monolayer spreads on heating as "solid" because the spreading begins at a temperature of ~64 °C; i.e., well below the melting point of the monoclinic phase of bulk C32 at 68 °C. To gain some insight into the structure of this particle, it is of interest to consider its volume as a function of temperature plotted in the left insert in Fig. 1 as determined from our AFM images [including those in Fig. 1(b)]. We see that there is a slight increase in its volume up to a temperature of ~61 °C at which point it decreases abruptly. This sudden drop in volume suggests that some structural change has occurred in the particle to a more volatile phase from which molecules can evaporate more readily into the interfacial gas phase.

Further evidence of a structural change occurring in the C32 films prior to the spreading of the perpendicular monolayer comes from grazing-incident-angle synchrotron x-ray diffraction experiments. Measurements were conducted at beam line 6ID-B of the Advanced Photon Source using an x-ray wavelength of 0.765 Å and a two-dimensional detector. The sample had about the same coverage and was prepared similarly to the one used in the AFM measurements of Fig. 1(b), including the initial heating cycle. In Fig. 2, we show scans of the x-ray intensity taken with the wave vector transfer **Q** aligned parallel to the $SiO_2$ surface. At each value of $Q$, the intensity has been integrated over a narrow range 0.02 Å$^{-1}$ to 0.04 Å$^{-1}$ in a direction normal to the surface in order to improve statistics but to exclude contributions from dynamical scattering near the C32 critical angle. Below a temperature of 56 °C, we see a diffraction pattern characteristic of a polycrystalline film consisting of three features: a dominant peak at $Q \sim 1.5$ Å$^{-1}$, a shoulder on its leading edge, and a broader peak at $Q \sim 1.66$ Å$^{-1}$. These features cannot be indexed by the bulk monoclinic structure reported for C32 at room temperature [15]. In particular, the small splitting of the shoulder and dominant peak suggest that they may originate from different phases. Therefore, we tentatively identify the two stronger peaks with a crystalline phase C' corresponding to an ensemble of 3D mesa-shaped C32 particles similar to the one in the AFM images of Fig. 1(b) and having a distribution of heights. Indexing the lower and upper of these peaks as the (110) and (200) reflections of an orthorhombic unit cell, respectively, we obtain lattice constants $a = 7.57$ Å and $b = 4.98$ Å in reasonable agreement with those found for orthorhombic particles of C32 growing on a Ag(111) surface [16].

As shown in Fig. 2, between 56 °C and 58 °C, we observe a transition from the double-peak feature near $Q \sim 1.5$ Å$^{-1}$ to a single broad peak and the disappearance of the weaker peak at $Q \sim 1.66$ Å$^{-1}$. These changes indicate a transition in the film to a phase characterized by higher symmetry and much shorter coherence length than in the low-temperature C' phase. Following previous suggestions of a rotator phase for interfacial alkane molecules of intermediate length [9,10], we interpret the changes in the x-ray scans in the range 56–58 °C as indicating a transition from the 3D crystalline phase C' to a 3D rotator (plastic) phase R' in which the C32 molecules are orientationally disordered about their long axis. Compared to the C' phase, the R' phase has much shorter range translational order and exchanges molecules more readily with the interfacial gas phase.

Using the AFM images as a guide, we tentatively identify the weak and broad diffraction peak at $Q = 1.49$ Å$^{-1}$ at the highest temperatures (65–68 °C) with the perpendicular monolayer phase [see Fig. 1(b)] and the weak shoulder at $Q = 1.50$ Å$^{-1}$ observed below 57 °C with a remnant of the metastable dragonfly phase in Fig. 1(a). These peak positions are close to the value of 1.51 Å$^{-1}$ found for the perpendicular monolayer phase that has hexagonal symmetry in the surface freezing effect of bulk C32 [17].

It is useful to summarize our results in an equilibrium phase diagram shown in Fig. 3 where we have plotted the C32 chemical potential $\mu$ measured with respect to that of the bulk liquid as a function of temperature. The solid black line segments labeled C', R', and L represent the chemical potential of the bulk crystalline, rotator, and liquid phases, respectively, where the prime indicates interfacial structures distinct from the freestanding bulk phases. We suggest that there is a first order transition from the 3D R'-phase to a perpendicular monolayer phase at a temperature $T^-\sim 64$ °C and from this phase to the 3D L-phase at $T^+\sim 69$ °C. The solid line segment between $T^-$ and $T^+$ represents the stability region of the perpendicular monolayer phase. We interpret the monolayer spreading behavior observed as the transition temperatures $T^-$ and $T^+$ are crossed in the direction toward the perpendicular monolayer phase to result from a slow relaxation to equilibrium. Similarly, we attribute the recession of the monolayer to a slow relaxation to the respective 3D phases as these phase boundaries are crossed in the opposite direction. Both the R' and perpendicular monolayer phases are characterized by shorter-range translational order than in the C' phase; but, as indicated by the slope of their stability lines in the phase diagram, have lower entropy than the nonwetting bulk liquid phase.

Preliminary AFM and synchrotron x-ray measurements indicate that a similar phase diagram applies to submonolayer C32 deposited from solution onto highly-oriented pyrolytic graphite as well as to C24, C30, and C36 deposited on the same $SiO_2$ surfaces that we have used in this study. These results raise a number of interesting questions: Why does the bulk liquid fail to wet the underlying parallel layers of molecules immediately adjacent to the $SiO_2$ surface and is the delayering transition of the perpendicular monolayer phase related to conformational changes in the molecules? What are the microscopic mechanisms driving the spreading and receding of the perpendicular monolayer phases? And at what (shorter) alkane chain length will a crossover to complete wetting occur as observed for pentane on graphite?

We thank D. S. Robinson for technical support at the 6-ID beam line of the Advanced Photon Source. This work was supported by the U.S. National Science Foundation under Grant No. DMR-0411748, by the Chilean government under FONDECYT Grant Nos. 1010548 and 701058, and by the U.S. Department of Energy under Contract Nos. W-7405-Eng-82 and W-31-109-Eng-38.

## References

*Permanent address: Technische Physik, Universität des Saarlandes, D 66041 Saarbrücken, Germany.

†Present address: Infineon Technologies SC300 GmbH & Co. KG, 01099 Dresden, Germany.

††Corresponding author.

**Figure captions**

Fig. 1. Topographic AFM images of a low-coverage C32 film taken in the noncontact mode. (a) At room temperature after deposition from solution; (b) temperature dependence of the images in the second heating cycle; and (c) images taken on cooling the sample after its second heating in (b). Left insert shows the temperature dependence of the particle volume in (b); right insert shows a height cross section taken along the line in the image at 66 °C in (b). All heights are measured relative to the one to two layers of C32 molecules that lie with their long-axis parallel and immediately adjacent to the $SiO_2$ surface.

Fig. 2. Temperature dependence of the in-plane synchrotron x-ray diffraction scans of a C32 film prepared similarly to that in Fig. 1. Intensities are calculated from the 2D detector after integrating over a limited $Q$ range in a direction perpendicular to the $SiO_2$ surface as described in the text.

Fig. 3. Proposed phase diagram for the submonolayer C32 film plotted in the $\mu$-$T$ plane where the chemical potential $\mu$ is measured relative to that of the bulk C32 liquid (L). C' and R' refer to the interfacial bulk C32 crystalline and rotator phases, respectively. The solid line segment between $T^-$ and $T^+$ denotes the stability region of the perpendicular monolayer phase.

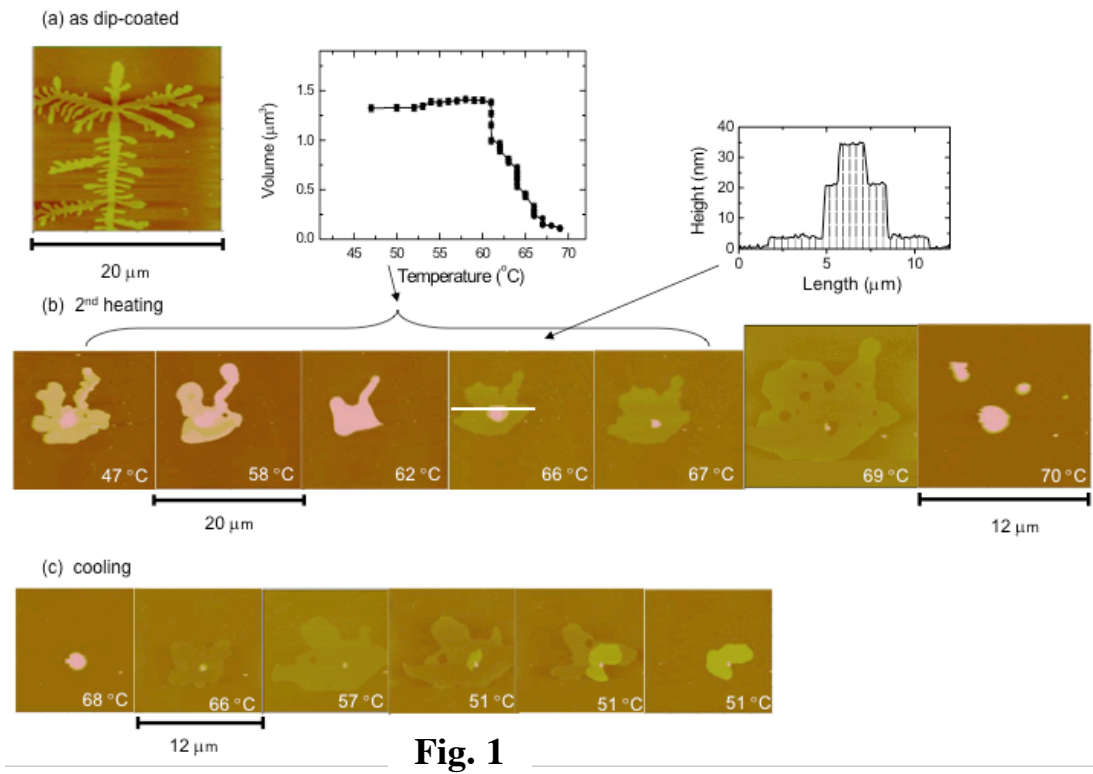

Fig. 1

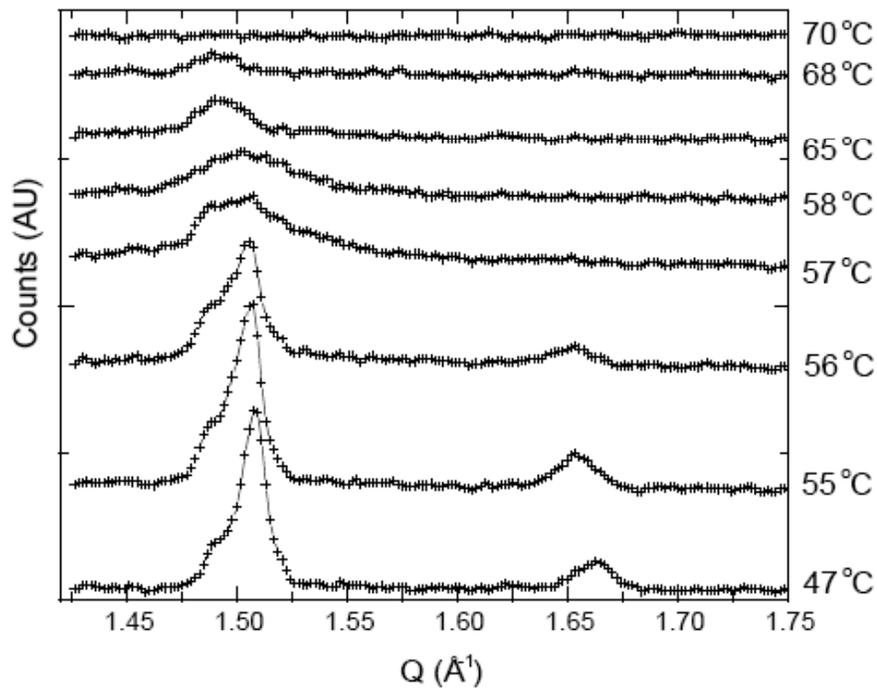

Fig. 2

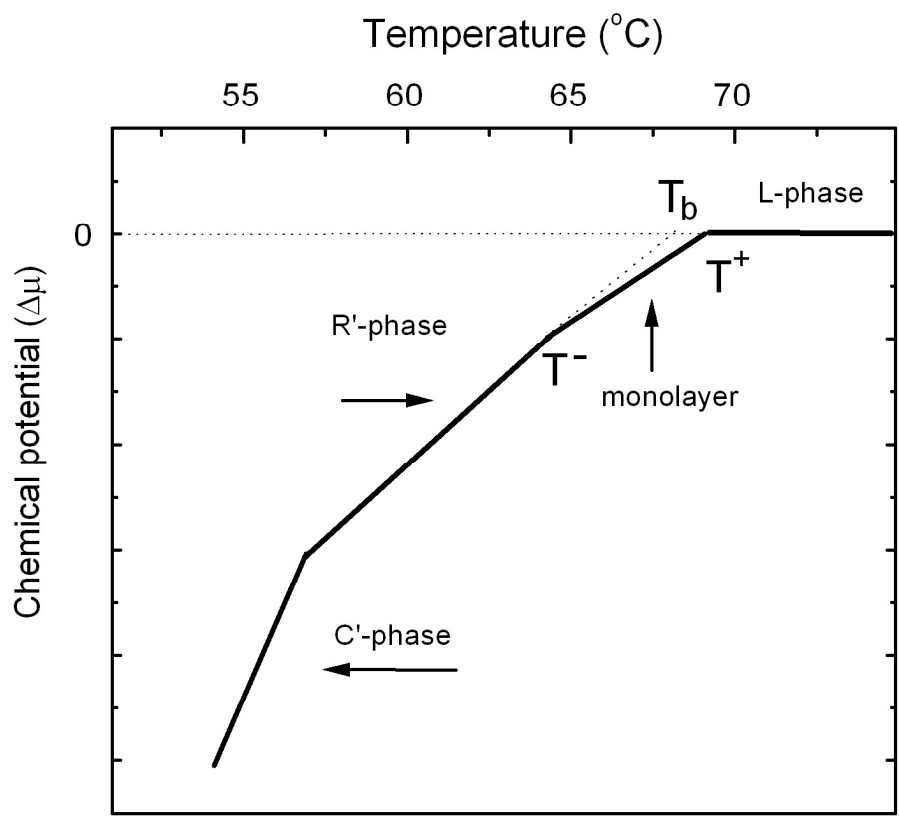

**Fig. 3**